\documentclass[pre, twocolumn, reprint, amssymb, amsmath, amsfonts, a4paper, aps, floatfix, superscriptaddress, showpacs, twoside]{revtex4-1}
\usepackage{tabularx, xspace, rotating, units, enumerate, ucs}

\newcommand{\Epi}{\affiliation{Department of Epileptology, University of Bonn, Sigmund-Freud-Stra{\ss}e~25, 53105~Bonn, Germany}}
\newcommand{\HISKP}{\affiliation{Helmholtz Institute for Radiation and Nuclear Physics, University of Bonn, Nussallee~14--16, 53115~Bonn, Germany}}
\newcommand{\IZKS}{\affiliation {Interdisciplinary Center for Complex Systems, University of Bonn, Br\"uhler Stra\ss{}e~7, 53175~Bonn, Germany}}
\newcommand{\ICBM}{\affiliation {Theoretical Physics/Complex Systems, ICBM, Carl von Ossietzky University of Oldenburg, \\Carl-von-Ossietzky-Stra\ss{}e~9--11, Box~2503, 26111~Oldenburg, Germany}}
\newcommand{\RNS}{\affiliation{Research Center Neurosensory Science, Carl von Ossietzky University of Oldenburg,\\ Carl-von-Ossietzky-Stra\ss{}e~9--11, 26111~Oldenburg, Germany}}
\newcommand{\Maryland}{\affiliation{Institute for Physical Science and Technology, University of Maryland, College Park, MD 20742-2431, U.S.A.}}

\newcommand{\abs}[1]{\left\lvert #1 \right\rvert}

\newcommand{\kl}[1]{\left( #1 \right)}
\newcommand{\klg}[1]{\left\{ #1 \right\}}
\newcommand{\kle}[1]{\left[ #1 \right]}

\newcommand{\defi}{\mathrel{\mathop:}=}

\begin{document}

\title{Extreme events in excitable systems and mechanisms of their generation}

\author{Gerrit Ansmann}
\Epi \HISKP \IZKS

\author{Rajat Karnatak}
\ICBM

\author{Klaus Lehnertz}
\Epi \HISKP \IZKS

\author{Ulrike Feudel}
\ICBM \RNS \Maryland

\begin{abstract}
We study deterministic systems, composed of excitable units of FitzHugh--Nagumo type, that are capable of self-generating and self-terminating strong deviations from their regular dynamics without the influence of noise or parameter change.
These deviations are rare, short-lasting, and recurrent and can therefore be regarded as extreme events.
Employing a range of methods we analyze dynamical properties of the systems, identifying features in the systems' dynamics that may qualify as precursors to extreme events.
We investigate these features and elucidate mechanisms that may be responsible for the generation of the extreme events.
\end{abstract}


\maketitle
\section{Introduction}

Many natural, technological, or social systems are capable of recurrently generating large impact events \cite{Bunde2002, Sornette2003, Albeverio2006}.
Well known are earthquakes, tsunamis or extreme weather events---such as heat waves, droughts, floods, heavy precipitation, or tornadoes---that can lead to disasters when interacting with exposed or vulnerable human or natural systems.
Other examples include epileptic seizures in the human brain~\cite{Lehnertz2006}, rogue waves in the ocean or in optical systems \cite{Pleskachevsky2012, Kharif2003, Solli2007}, harmful algal blooms in marine ecosystems~\cite{Anderson2012b}, large-scale blackouts in power supply networks~\cite{Dobson2007}, market crashes~\cite{Feigenbaum2001}, mass panics~\cite{Helbing2001}, or wars~\cite{Hobsbawm1994}.

The temporal evolution of a relevant observable of such systems usually exhibits small-scale fluctuations around some well-defined level (e.g., derived from the long-term average of available data).
Occasionally, however, this observable shows abrupt excursions to values that differ significantly from this level.
Such rare and recurring emergences of unusually large or small values are of paramount importance since they can indicate extreme events.

Though essential influencing factors are known for many of the aforementioned events, the exact mechanism of their emergence is not well understood.
Which of those factors or which combination of them is the main trigger to start the development of an extreme event, is often an open question and therefore subject of current research.
Besides the need of an improved understanding of the generation and termination of extreme events, there is an urgent call for their predictions.
Progress along these lines may be achieved through the analysis of time series of system observables \cite{Sornette2002, Altmann2005, Bunde2005, Hallerberg2007, Mormann2007, Dakos2008, Manshour2009, Scheffer2009, Kramer2012b, Boettiger2013} or through the development and investigation of models.
Many of the aforementioned systems can be modeled as excitable systems or as composed of excitable units \cite{Meron1992, Cross1993, Cartwright1997, Sagues2007}.
In general, excitability refers to the system's capability to develop a large pulse of activity in response to some endogenous or exogenous triggering mechanism~\cite{Izhikevich2007}.
This pulse lasts for some time and, depending on conditions, may either stop or propagate through space.

One of the most simple and widely studied excitable systems is the FitzHugh--Nagumo system (also known as Bonhoeffer--van der Pol model) \cite{VanDerPol1928, Bonhoeffer1948, FitzHugh1961, Nagumo1962}, which captures the qualitative essence of neuronal firing through a simple algebraic form of the evolution equations~\cite{Rocsoreanu2000}.
It is widely used as a model for excitable behavior in neural and cardiac nonlinear activities \cite{Koch1999, Glass1991}.
Phenomena observed in FitzHugh--Nagumo systems include pattern formation \cite{Winfree1991, Jung2000, Perc2005}, firing death \cite{Hennig2007, Ciszak2013}, noise-induced phenomena \cite{Jung1998b, Toral2003, Acebron2004, Zaks2005, Patidar2009, Zambrano2010}, diversity-induced oscillations \cite{Cartwright2000, Vragovic2006}, and aspects of synchronization \cite{Hennig2007, Hennig2008}.

Here, we address the question, whether and how excitable systems of FitzHugh--Nagumo units are capable of generating extreme events, which would allow to study the underlying generating mechanisms.
We report on deterministic model systems of coupled FitzHugh--Nagumo units that are capable of generating extreme events and analyze the mechanisms behind them.
This paper is organized as follows.
In Sec.~\ref{systems} we describe our model systems and report on the extreme events they exhibit and their properties.
In Sec.~\ref{revlyap} we propose a method to detect precursors to extreme events in model systems in general and apply them to our model systems.
For those systems we then present our findings on the mechanisms behind the events (Secs.~\ref{gentwo} and~\ref{genmany}).
Finally, in Sec.~\ref{conclusions} we draw our conclusions.

\section{Model systems and their behavior}\label{systems}
We consider systems of $n$~diffusively coupled FitzHugh--Nagumo units ($i \in \klg{1, \ldots, n}$) whose $i$\nobreakdash-th unit is described by the following differential equations:
	\begin{align}
		\dot{x}_i & = x_i (a_i-x_i) (x_i-1) - y_i + k \sum\limits_{j=1}^{n} A_{ij} (x_j - x_i) \notag\\
		\dot{y}_i & = b_i x_i - c_i y_i
		\label{eq:systems}
	\end{align}
Here $a_i$, $b_i$ and $c_i$ are internal parameters of the unit, $k$~is the coupling strength, and $A \in \klg{0,1}^{n \times n}$ is the symmetric adjacency matrix ($A_{ij} = A_{ji} = 1$, iff units $i$ and~$j$ are coupled).

In particular, we regard the following three systems:
\begin{enumerate}[(A)]
	\item \label{twounits} A system of $n=2$ mutually coupled units, i.e., $A = \kl{\begin{smallmatrix}0&1\\ 1&0\end{smallmatrix}}$.
	The parameters $a$ and~$c$ are identical for both units: $a_1=a_2=a=-0.025794$ and $c_1=c_2=c=0.02$; the parameter~$b$ is mismatched: $b_1=0.0065$, $b_2=0.0135$.
	The coupling strength~$k$ is~0.128.
	\item \label{manyunits} A system of $n=101$ completely coupled units, i.e., $A_{ij} = 1\,\forall\,i,j$.
	The parameters $a$ and~$c$ are identical for all units: $a_i=a=-0.02651\,\forall\,i$ and $c_i=c=0.02\,\forall\,i$; the parameter~$b$ is mismatched: $b_i = 0.006 + \tfrac{i-1}{n-1}\cdot 0.008$ ($\Rightarrow 0.006 \leq b_i \leq 0.014 \,\forall\,i$).
	The coupling strength~$k$ is 0.00128.
	\item \label{smallworld} A system of $n=10000$ units coupled with a small-world topology~\cite{Watts1998}:
	The units are arranged on a $100 \times 100$ lattice with cyclic boundary conditions (torus) and each unit is connected to its 60~nearest neighbors.
	Then each edge is rewired with a probability of 0.2, i.e., it is removed and replaced by an edge between two randomly chosen units.
	The parameters $a$ and~$c$ are identical for all units: $a_i=a=-0.0276\,\forall\,i$ and $c_i=c=0.02\,\forall\,i$; the parameter~$b_i$ is randomly drawn from the uniform distribution on $\kle{0.006, 0.014}$ for each unit.
	The coupling strength $k$ is $0.0021\bar{3}$.
\end{enumerate}
Note that for each system the product of coupling strength and average degree (i.e., the average number of units to which a given unit is connected) is 0.128 and the values or ranges, respectively, for the control parameters $a$, $b$, and~$c$ are comparable, if not identical.
The slight variation of the parameter $a$ is to align the dynamical behaviors of the systems.
While inhomogeneous units in general are more realistic, it is for simplicity's sake that we choose to make the units inhomogeneous in $b$ and select to distribute the $b_i$ uniformly for systems \ref{manyunits} and~\ref{smallworld}.

If any of these units are uncoupled, $x_i\kl{t}$ exhibits oscillations with peak amplitudes between 0.88 and 0.95 and $y_i\kl{t}$ exhibits oscillations with peak amplitude between 0.17 and 0.21.
These oscillations correspond to relaxation oscillations and have a period length between 106 and 193, depending on parameters.
If the units are coupled, they exhibit a chaotic behavior, as implied by the Lyapunov exponents~\cite{Benettin1980}:
We obtain $\Lambda_1=0.0071$, $\Lambda_2=0.0000$, $\Lambda_3=-0.0512$, $\Lambda_4=-0.1870$ for System~\ref{twounits} and $\Lambda_1=0.0053$, $\Lambda_2=0.0000$, $\Lambda_3=-0.0186$, $\Lambda_4=-0.0197$ as the four largest Lyapunov exponents for System~\ref{manyunits}.
For the temporal evolution of the average of $x_i$ over all units of System~\ref{smallworld}, we observe a broadband power spectrum, which indicates this system to also be chaotic.

System~\ref{twounits} was integrated using a 4\textsuperscript{th}-order Runge--Kutta method with a time step of $0.01$.
Systems \ref{manyunits} and~\ref{smallworld} were realized with Conedy, a tool to compute arbitrary dynamics with arbitrary coupling topologies~\cite{Rothkegel2012}.
The dynamics was integrated with the Runge--Kutta--Fehlberg procedure, whose step size was adapted such that the estimated relative error did not exceed $10^{-5}$.
We tried other integration schemes and other numerical precisions but could not observe an influence on the systems' dynamical behavior.
Also, the same qualitative dynamical behavior could be observed for a range of control parameters, even though a relatively small one (we will report on the parameter dependence of the observed phenomena elsewhere~\cite{Karnatak2014}).
For each of the following observations and analyses at least 10000 initial time units were discarded.
The choice of the initial conditions (near the attractor) had no influence on our observations.

\begin{figure*}
\includegraphics{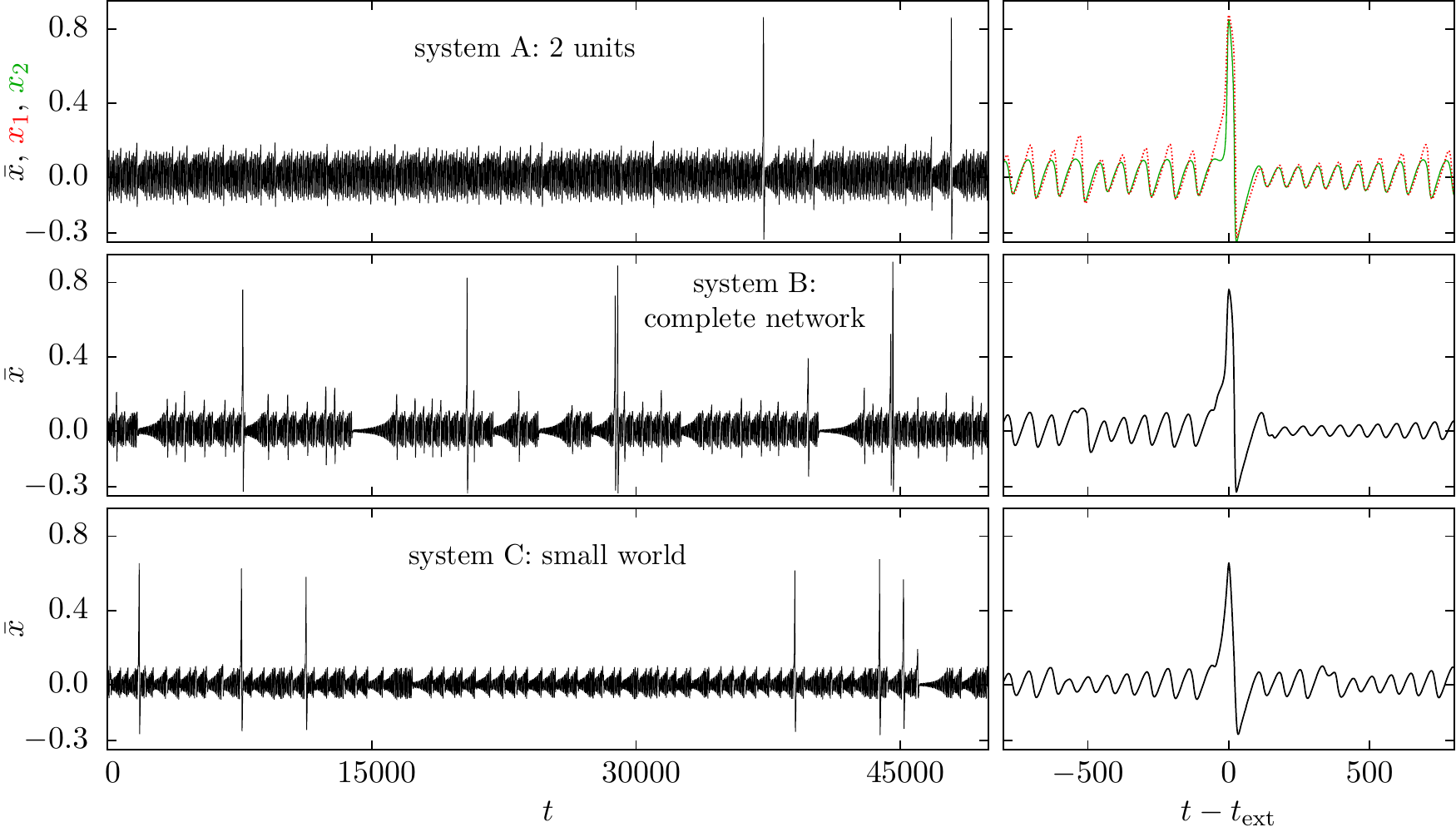}
	\caption{(Color online) (left) Exemplary temporal evolutions of~$\bar{x}$ for the investigated systems.
	(right) Excerpt centered around the first extreme event (at $t_\text{ext}$) of the respective time series.
	For System~\ref{twounits} (first row) the individual time series of $x_1$ (red, dotted line) and $x_2$ (green, solid line) are shown.}
	\label{fig:timeseries}
\end{figure*}

In Fig.~\ref{fig:timeseries} (left panels), we show typical time series of the average value of the first dynamical variable, $\bar{x}\kl{t} = \tfrac{1}{n} \sum_{i=1}^n x_i\kl{t}$ (which we use as the main observable in the following) for all three considered systems.
Predominantly, $\bar{x}\kl{t}$ exhibits oscillations with small amplitude ($-0.2< \bar{x}\kl{t}<0.3$) and without any apparent regularity, which we are going to refer to as \textit{low-amplitude oscillations.}
The period length of these oscillations---estimated as the distance between adjacent local maxima---varies around~75 (System~\ref{twounits}: $80\pm7$, System~\ref{manyunits}: $71\pm12$, System~\ref{smallworld}: $69\pm8$).
However, sometimes we observe events at which $\bar{x}\kl{t}$ exhibits amplitudes that are at least six times higher than the amplitudes of the low-amplitude oscillations.
These events qualify as extreme events in our understanding since the observable~$\bar{x}$ exhibits unusually large values; the events are rare in comparison to the usual timescales of the system dynamics (low-amplitude oscillations), and they are recurring.
The right panels of Fig.~\ref{fig:timeseries} show a close-up view of one extreme event for each system.
Particularly, for System~\ref{twounits}, we illustrate the behavior of both units separately.
Looking at the time series, it appears that unit~1 ($b_1=0.0065$) recruits unit~2 ($b_2=0.0135$) towards the extreme event.

\begin{figure}
	\includegraphics{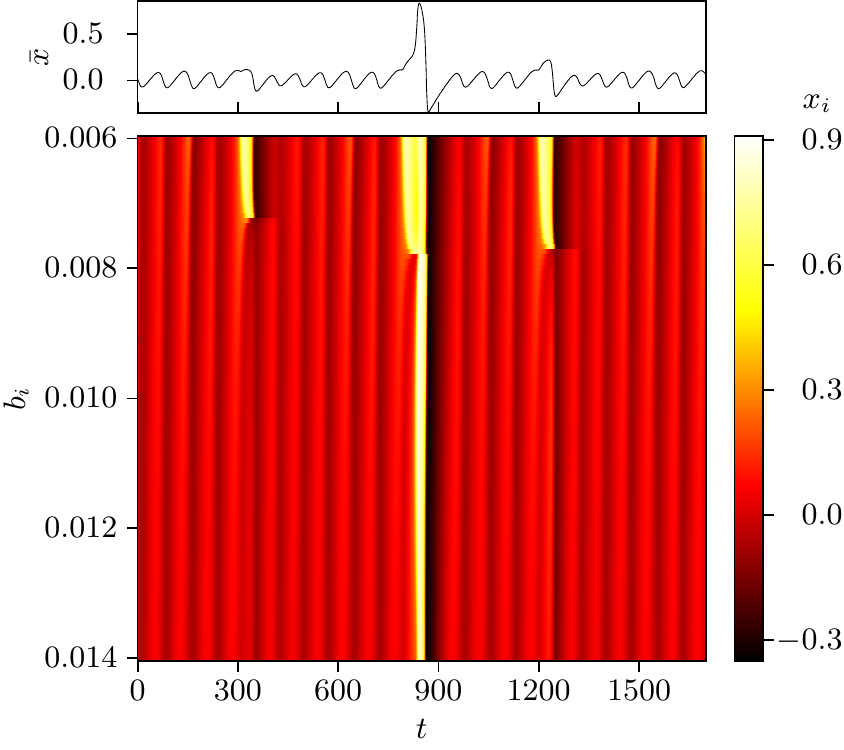}
	\caption{(Color online) Temporal evolutions of $x_i$ for each unit of System~\ref{manyunits} around an extreme event.
̂	The units are indexed by their value of $b_i$.
	For reference, the temporal evolution of $\bar{x}$ is shown on top.}
	\label{fig:manyunits}
\end{figure}

In Fig.~\ref{fig:manyunits} we show the normal as well as the extreme behavior of System~\ref{manyunits} in detail:
At $t\approx800$ we observe an extreme event:
First, units with lowest~$b_i$ become \textit{excited} (i.e., the corresponding~$x_i$ assumes a high value), an event which we are going to refer to as \textit{proto-event} in the following.
Shortly afterwards this \textit{cluster} of excited units seems to recruit the remaining ones, causing all units to be excited simultaneously, which constitutes the extreme event.
At $t \approx 300$ and $t \approx 1200$ we also observe proto-events without an extreme event following.
After these excitations, the cluster lasts for about one low amplitude oscillation (note the clear-cut ``shadow'').
The average time between such \textit{proto-events} is 1520 with a standard deviation of 1331, while the average time between the extreme events is 7785.
During the rest of the time, all units exhibit low-amplitude oscillations and a high level of synchrony.
For this system we also observe \textit{double extreme events,} which consist of two subsequent events, the first of which is smaller in amplitude and the only of the two that is preceded by a proto-event.

\begin{figure*}
	\includegraphics{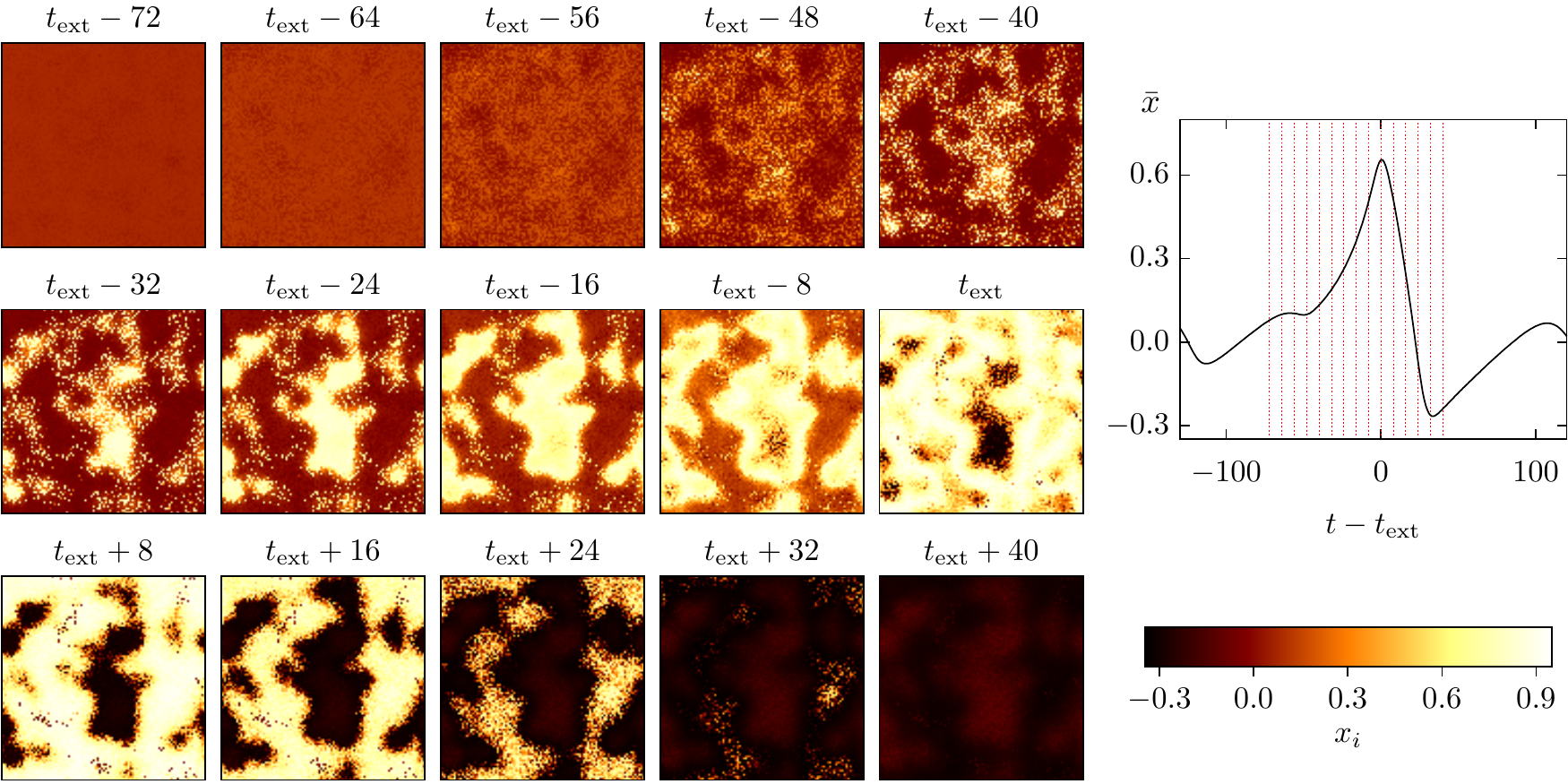}
	\caption{(Color online) (left) Snapshots of the spatial distribution of $x_i\kl{t}$ for one realization of System~\ref{smallworld} around an extreme event (at $t_\text{ext}$).
	Units are represented by pixels, which are arranged according to the lattice underlying the small-world network.
	The respective value of the dynamical variable $x$ is color-coded.
	(top right) Temporal evolution of $\bar{x}$, with snapshot times shown by red vertical lines.
	The event is the first one in Fig.~\ref{fig:timeseries}, third row.}
	\label{fig:smallworld}
\end{figure*}

Fig.~\ref{fig:smallworld} shows an extreme event at $t_\text{ext}$ for System~\ref{smallworld} in detail.
At $t \approx t_\text{ext} - 40$ we observe that excited units form a few localized clusters, which grow with increasing speed until $t_\text{ext}$.
At this time most units are excited, the only exception being some units within some of the initially excited clusters.
These units have already become refractory.
Concordantly, we observe the height of an extreme event in terms of the amplitude of $\bar{x}$ to be lowest for System~\ref{smallworld} (see Fig.~\ref{fig:timeseries}).

For all three systems, there is one general feature of the extreme events:
One unit or a group of units become(s) excited and recruit(s) the remaining one or ones to form an extreme event.

\begin{figure}
	\includegraphics{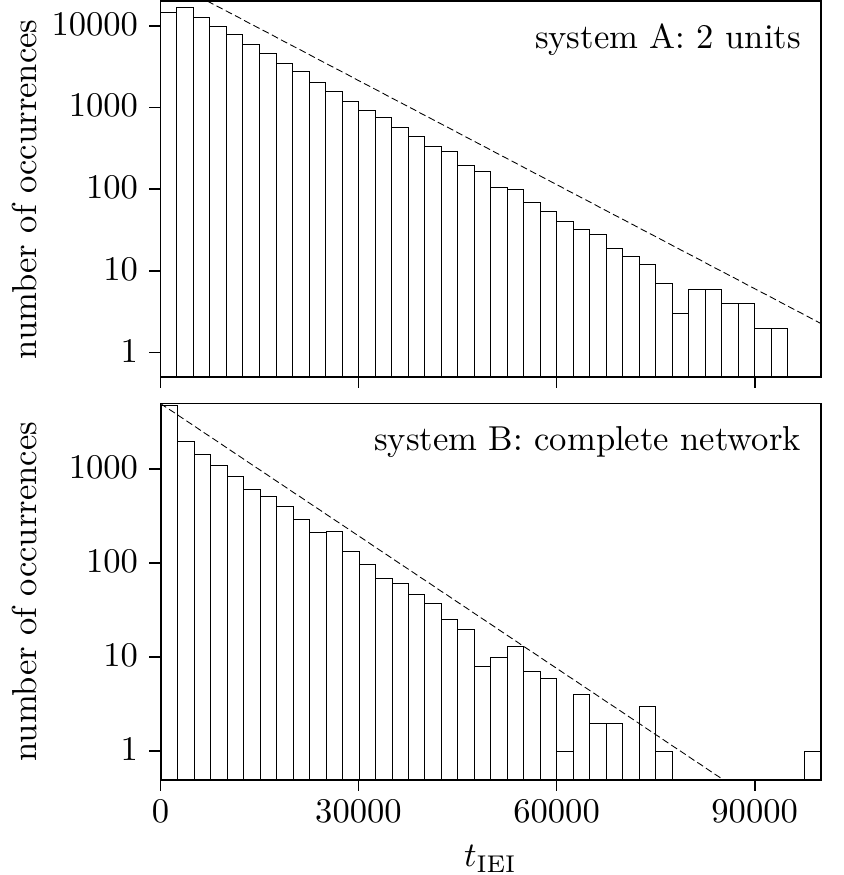}
	\caption{(left) Histogram of the inter-event intervals $t_\text{IEI}$ for two of the investigated systems.
	The observation time was $9\cdot10^9$ for System~\ref{twounits} and $10^8$ for System~\ref{manyunits}.
	The dashed lines are multiples of $\exp\kl{- r t_\text{IEI}}$.
	Here, $r=9.8 \cdot 10^{-5}$ (System~\ref{twounits}) and $r=1.0 \cdot 10^{-4}$ (System~\ref{manyunits}), respectively, are the rates of an exponential distribution fitted to the data (for $t_\text{IEI}>200$).}
	\label{fig:statistics}
\end{figure}

We restrict ourselves to systems \ref{twounits} and~\ref{manyunits} for the following analyses, since System~\ref{smallworld} is not feasible for these.
This is due to the randomness involved in its creation and the fact that only realization-specific results are obtainable for this system, and averaging of these would be meaningless, given our methods of analysis.
Also, for the purposes of automatizing the analyses, we define an extreme event to be an interval of time with $\bar{x}\kl{t}>0.6$, and we define the time of such an event to be the start of the corresponding interval.
Since there are very few local maxima of $\bar{x}$ with $0.3<\bar{x}\kl{t}<0.6$ and the extreme events are very homogeneous in form, we consider a more sophisticated event detection unnecessary for our purposes.

In Fig.~\ref{fig:statistics} we show the estimated distributions of inter-event intervals for systems \ref{twounits} and~\ref{manyunits}.
We observe both distributions to be nearly exponential, which would be the result for a Poissonian process.

\section{Searching for precursors}\label{revlyap}
In this section, we propose and apply a method to investigate, whether and when generating mechanisms of extreme events come into action in our systems.
This in turn may be reflected by precursors.
To this end, we require some \textit{generating mechanism} to cause an extreme event inevitably or at least with a high probability.
Also, we assume that such a mechanism is reflected by the dynamical variables and is (in general) not disabled by small perturbations to these variables.

\subsection{Method}
To detect such a generating mechanism, we perform the following \textit{perturbation analysis:}
Given a trajectory $\kl{\boldsymbol{x}\kl{t},\boldsymbol{y}\kl{t}}$ leading to an extreme event at $t_\text{ext}$, we estimate the probability $q\kl{t_\text{per}, \varepsilon}$ that the event cannot be observed anymore if we perturb the system at some $t_\text{per}<t_\text{ext}$ with amplitude~$\varepsilon$.

Generally, we expect $q$ to tendentially increase with increasing~$\varepsilon$ and decrease with increasing $t_\text{per}$, given a chaotic dynamical system that self-generates rare events and all of whose dynamical variables are perturbed in a reasonable way.
We focus on $q$\nobreakdash-isolines in the $\varepsilon$--$t_\text{per}$ plane and, to simplify the description, treat them as functions of $t_\text{per}$.
As such they can be regarded as indicators of the ``sensitivity'' of the extreme event to perturbations at a given $t_\text{per}$.
Therefore the $q$\nobreakdash-isolines should also indicate a generating mechanism coming into action by a strong increase, since such a mechanism should make it much harder to ``prevent'' the extreme event.
We thus look for such increases of the $q$\nobreakdash-isolines which exceed the generally expected increase due to the chaoticity of the system, as quantified by the maximum Lyapunov exponent.
The advantage of this approach over regarding the maximum Lyapunov exponent is that it specifically considers the effect of finite perturbations on the occurrence of extreme events.
This way, we can detect the existence of a precursor or generating mechanism, respectively, however, we cannot obtain any further information about its nature.

To perturb the system's state $\kl{\boldsymbol{x}\kl{t_\text{per}},\boldsymbol{y}\kl{t_\text{per}}}$, we here generated two $n$-dimensional vectors $\boldsymbol{z}$ and~$\boldsymbol{w}$ with unit length and random direction and added $\varepsilon \boldsymbol{z}$ to $\boldsymbol{x}\kl{t_\text{per}}$ and $\eta \varepsilon \boldsymbol{w}$ to $\boldsymbol{y}\kl{t_\text{per}}$.
The factor~$\eta$ accounts for the fact that the attractor of the individual units extends differently in $x$\nobreakdash- and $y$\nobreakdash-direction.
To estimate $q\kl{t_\text{per}, \varepsilon}$ for a given investigated $\kl{t_\text{per}, \varepsilon}$, we employed a number of realizations of the perturbations (i.e., different directions of $\boldsymbol{z}$ and~$\boldsymbol{w}$), calculating the ratio of cases, for which no extreme event happened within a certain allowance around $t_\text{ext}$.
Note that we chose not to average over different events since this would require a~priori that generating mechanisms always come into play at about the same time in relation to an extreme event.

\begin{figure*}
	\includegraphics{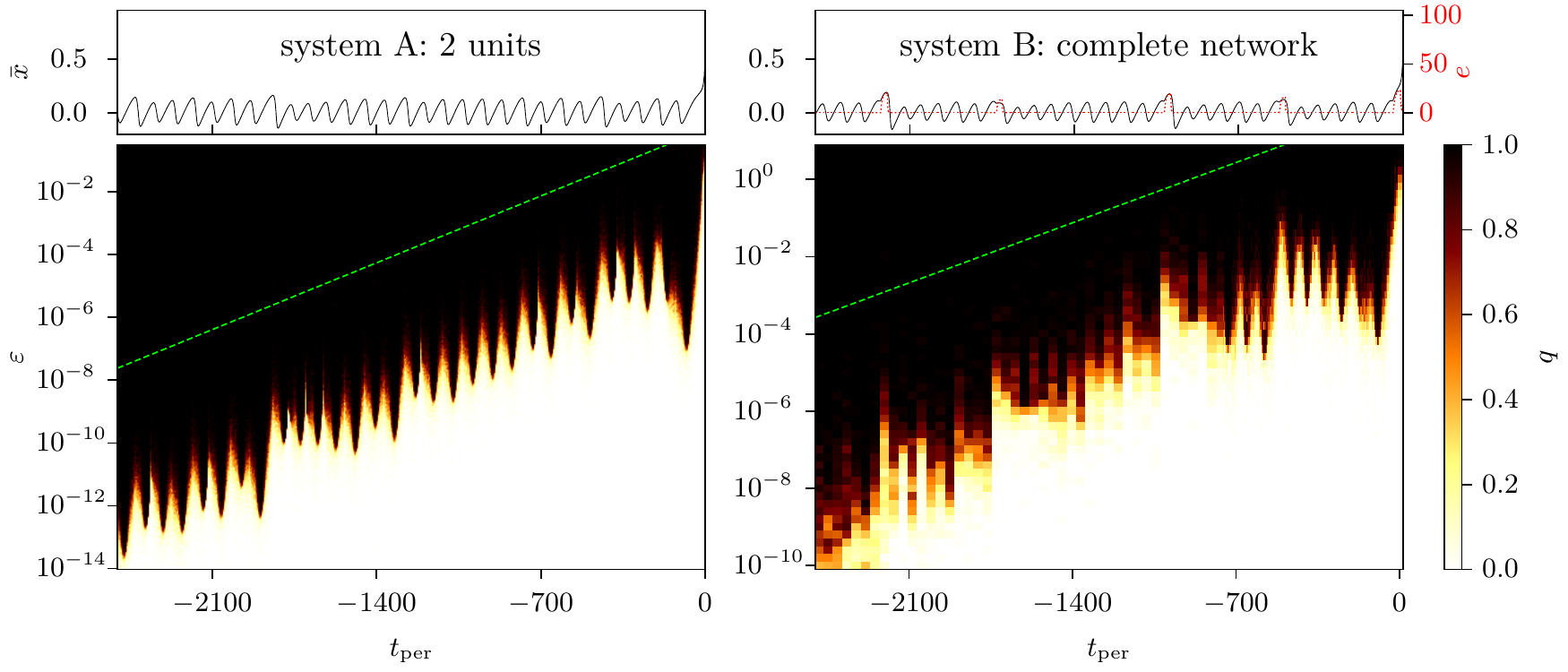}

	\caption{(Color online) Results of the perturbation analysis for an exemplary extreme event (at $t_\text{ext}=0$) of System~\ref{twounits} (left) and System~\ref{manyunits} (right), taking into account events that happen within 1.0 time units of $t_\text{ext}$ in the perturbed systems.
	(top) Temporal evolution of $\bar{x}$ for the unperturbed system.
	(bottom) Relative frequencies $q\kl{t_\text{per}, \varepsilon}$ that the shown event is ``prevented'' by a perturbation of amplitude~$\varepsilon$ at $t_\text{per}$, estimated using 64~realizations (i.e., different perturbations).
	For reference, we show $\exp\kl{\Lambda_1 t_\text{per}}$ (green, dashed line), with $\Lambda_1$ being the largest Lyapunov exponent of the system.
	For System~\ref{manyunits}, the time series of the number of excited units $e\kl{t} \defi \abs{\klg{i\middle|x_i\kl{t}>0.6}}$ is shown as additional reference.
}
	\label{fig:revlyap}
\end{figure*}

\subsection{Results}
\label{revlyapresults}

In the left part of Fig.~\ref{fig:revlyap} we show the result of such a perturbation analysis for an exemplary event of System~\ref{twounits}, taking into account events that happen within 1.0 time units of $t_\text{ext}$ in the perturbed system.
For a given $t_\text{per}$, we observe~$q$ to monotonically increase with~$\varepsilon$ from $q\approx0$ to $q\approx1$ over one order of magnitude of $\varepsilon$.
For a given $\varepsilon$, $q$ also decreases with $t_\text{per}$ as a general tendency, however, the transition between $q \approx 1$ and $q \approx 0$ is more intricate, alternating between high and low values of~$q$.
These alterations correspond to the low-amplitude oscillations of the unperturbed system, and reflect the dependence of the system's sensitivity to perturbations on the current phase of the low-amplitude oscillations.
Apart from these oscillations, the $q$\nobreakdash-isolines can be described quite well by $\alpha\exp\kl{\Lambda_1 t_\text{per}}$ for some $\alpha \in \mathbb{R}$, with $\Lambda_1$ being the largest Lyapunov exponent of the system.
However, we observe two cases of the $q$\nobreakdash-isolines increasing faster than expected considering $\Lambda_1$ and the influence of the low-amplitude oscillations:
one at $t_\text{per} \approx t_\text{ext}-1800$ and one at $t_\text{per} \approx t_\text{ext}-50$.
After the first incidence, however, the $q$\nobreakdash-isolines cease to increase until the effect of the initial increase is compensated, and therefore we do not consider it to be indicative of a generating mechanism.
The second case is a sharp increase of the $q$\nobreakdash-isolines, which is directly followed by the event.
This may indicate that a generating mechanism for the extreme events comes into action only at this time.

Considering only events in the perturbed system that happen within 1.0 time units of $t_\text{ext}$ only accounts for the case that the generating mechanism causes the event very shortly after coming into action in the unperturbed system.
Hence, it neglects the case that the generating mechanism might last longer and might cause an event after some delay.
If we do, however, consider this by also counting events that happen some time after $t_\text{ext}$, this does not affect the results qualitatively, but only lowers the maximum value obtained by $q$ slightly, which is most probably due to events that are unrelated to the ``original'' event.
Similar results are obtained if we count all events between the perturbation and $t_\text{ext}$ when estimating $q$.
In this case, for every $t_\text{per}$, $q$~decreases again at $\varepsilon \approx 10^{-1}$, i.e., for very large perturbations.
This is due to the perturbations exciting the system and thus causing extreme events by themselves or even being detected as extreme events themselves.

In the right part of Fig.~\ref{fig:revlyap} we show the result of the perturbation analysis for an exemplary event of System~\ref{manyunits}.
The results are mainly identical to those for System~\ref{twounits}, however there are faster increases of the $q$\nobreakdash-isolines, more precisely at $t\approx t_\text{ext} -2200$, at $t\approx t_\text{ext} -1700$, at $t\approx t_\text{ext} -1000$ and at $t\approx t_\text{ext} -500$.
These increases do however coincide with proto-events, before which a higher sensitivity of the system to perturbations was expected to some extent (since, e.g., the excited unit with highest $b_i$ in the unperturbed system is not excited in the perturbed system).
Furthermore, the effects of those increases are compensated later and we therefore do not consider them to be indicative of a generating mechanism.
Hence, we consider a generating mechanism only to come into action shortly (about 50~time units) before the extreme event, as already concluded for System~\ref{twounits}.

Both results shown here are exemplary for extreme events of the respective systems in all the aspects described above except for the specific position of incidents where the $q$\nobreakdash-isolines increase faster than expected considering $\Lambda_1$ (and of the proto-events).
For both systems, we therefore expect generating mechanisms only to come into play shortly (i.e., less than one low-amplitude oscillation) before the extreme events for both, system \ref{twounits} and~\ref{manyunits}.
We therefore focus on this interval, when further investigating extreme-event generating mechanisms in the following sections.
For System~\ref{manyunits} we more specifically focus on the proto-events, which are located in this interval.

\section{How an extreme event is generated in System~\ref{twounits}}\label{gentwo}

\subsection{Imperfect phase synchronization}
\begin{figure}
	\includegraphics{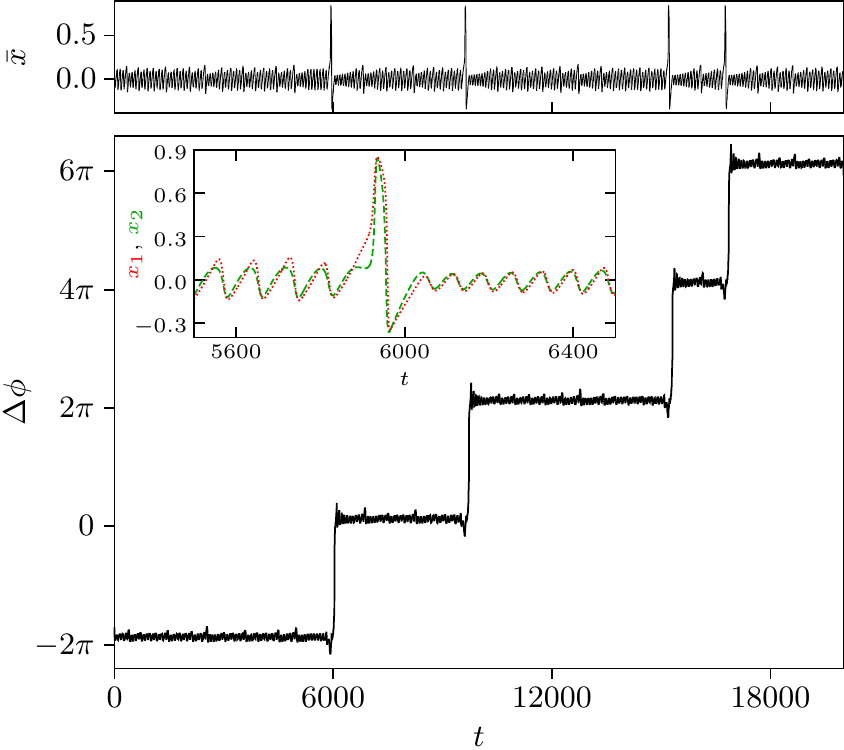}
	\caption{(Color online) (top) Temporal evolution of $\bar{x}$ for System~\ref{twounits}.
	(bottom) Estimated phase difference $\Delta\phi\kl{t}$ between the two units.
	(inset) Detailed behavior of the system around one example of the phase slips.}
	\label{fig:phaseslips}
\end{figure}

The top right part of Fig.~\ref{fig:timeseries} shows the behavior of the individual variables $x_1$ and $x_2$ for System~\ref{twounits} around an extreme event.
We observe that the trajectories of the units are phase-synchronized~(PS) and exhibit an apparent loss of this behavior at extreme events.
To quantify this and the characteristics of phase synchronization in general, we calculate the phase difference between the units.
To this end we employ the analytical signal approach \cite{Panter1965, Rosenblum1996}, based on the Hilbert transform of a system variable.
We construct analytical signals for System~\ref{twounits} using the variables $x_1$ and $x_2$ and calculate the instantaneous phases as $\phi_1(t)$ and $\phi_2(t)$ for units 1 and~2, respectively.
Consequently, the phase difference is defined as $\Delta\phi(t) \defi \phi_1(t)-\phi_2(t)$ (considering $1:1$~PS).
We also estimated the phases by interpolating between consecutive Poincar\'e surface crossings~\cite{Rice1944}, obtaining similar results.

Fig.~\ref{fig:phaseslips} shows the temporal evolution of the phase difference $\Delta \phi\kl{t}$.
We observe each extreme event to be associated with a phase slip, with unit~1 leading unit~2 by $2\pi$.
In coupled chaotic systems, such an interruption of the synchronous behavior by phase difference slips in multiples of $2\pi$ is referred to as imperfect phase synchronization (IPS) \cite{Zaks1999, Park1999, Blasius2000}.
Studies have related this behavior to the presence of a broad range of characteristic time scales in the dynamical system \cite{Zaks1999, Park1999}.
This can occur in a chaotic dynamical system when the chaotic attractor contains fixed points of saddle type; for example in the Lorenz system, where the saddle at the origin belongs to the closure of the chaotic set~\cite{Park1999}.
We expect a similar behavior in our system, and in the following subsection, we will look more closely at the dynamics of the system along with the properties of the saddle-type equilibrium at the origin.

\subsection{Role of the saddle-type equilibrium}
System~\ref{twounits} has a trivial equilibrium at the origin $\kl{x_1^*, y_1^*, x_2^*, y_2^*} \defi \kl{0,0,0,0}$, which interestingly is a saddle focus with two-dimensional stable and unstable manifolds, as seen from the eigenvalues of the Jacobian (see Appendix~\ref{sec:app1}).
The saddle quantity for the origin is $\sigma \approx -0.098394 <0$, which suggests that the origin is a simple saddle focus and does not lead to complicated dynamical scenarios, which could exist for $\sigma>0$ \cite{Shilnikov1965, Shilnikov1970}.
With the earlier observation regarding the Lorenz system \cite{Zaks1999, Park1999} in mind, we look at the role of the saddle focus behind the IPS and the extreme events for our case.
As it turns out, the saddle focus at the origin and its stable and unstable manifolds indeed play an important role in the generation of the extreme event.

\newcommand{\Dsmin}{\ensuremath{D_\text{s}}\xspace}
\newcommand{\Dumin}{\ensuremath{D_\text{u}}\xspace}

\begin{figure*}
	\includegraphics{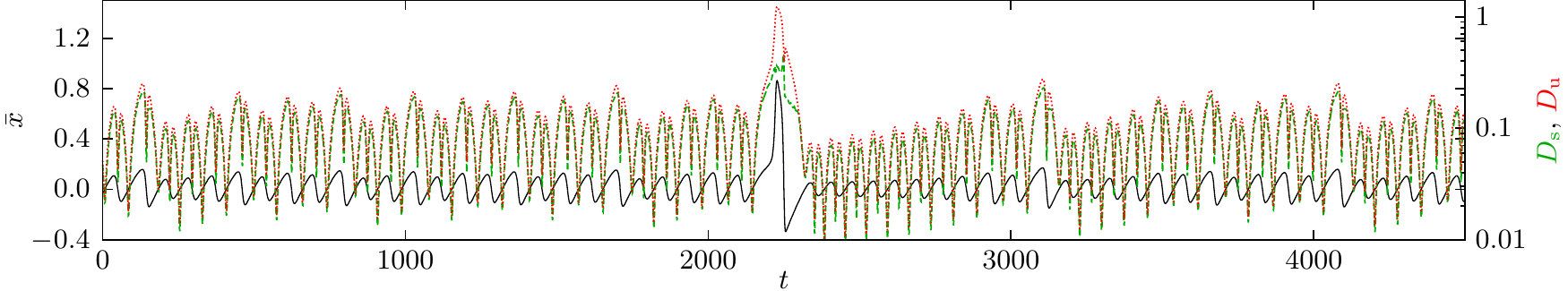}
	\caption{(Color online) Temporal evolutions of $\bar{x}$ (solid, black line) for System~\ref{twounits} and, on a logarithmic scale, of the distances \Dsmin (green, dashed line) and \Dumin (red, dotted line) of the system's state from the stable or unstable manifold, respectively.}
	\label{fig:manifolds}
\end{figure*}

\begin{figure}
	\includegraphics{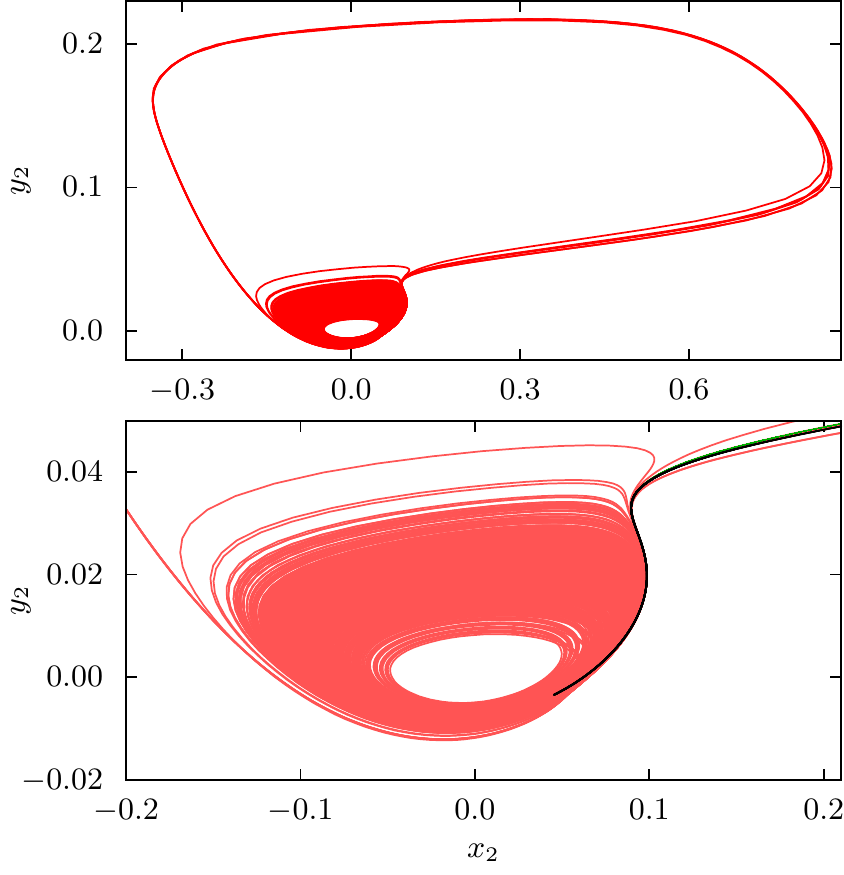}
	\caption{(Color online) Projection of the attractor of System~\ref{twounits} on the $x_2$--$y_2$-plane (red, dotted line).
	In the bottom zoom, two extreme-event trajectories are shown in black and green thick solid lines, illustrating the channel-like structure leading to the extreme event.}
	\label{fig:channel} \end{figure}

We calculated an approximation of the manifolds (see Appendix~\ref{sec:app1} for details) and the minimum Euclidean distances of the system's state to the stable and the unstable manifold, which we denote by \Dsmin and \Dumin, respectively.
In Fig.~\ref{fig:manifolds} we show the temporal evolutions of these distances.
We observe that, along with the state of the system in state space, the corresponding distances \Dsmin and \Dumin show similar oscillatory behavior.
This suggests that the trajectory of the system comes close and then departs away from the manifolds during the low-amplitude oscillations of~$\bar{x}\kl{t}$.
In case of an extreme event (e.g., for $t \approx 2200$ in Fig.~\ref{fig:manifolds}), however, the system exhibits a long excursion in state space (as shown in the top part of Fig.~\ref{fig:channel}) and gets farthest from the manifolds.

The appearance of these long excursions is closely related to the alignment of the manifolds.
In the parameter range where the emergence of extreme events is possible, these manifolds are located in such a way that there exists a small channel-like structure in state space through which the trajectory can escape for the long round trip (see Fig.~\ref{fig:channel}).
Since the arrangement of the manifolds in state space depends on the system parameters, this channel-like structure can open and close, either permitting or preventing the emergence of extreme events~\cite{Karnatak2014}.
Hence, the existence of the channel-like structures is the backbone of generating extreme events in System~\ref{twounits}.
On the other hand, the apparent randomness of the events in the system can be explained by the chaotic dynamics of the system, which causes the trajectories to enter the channel recurrently but aperiodically.
Because of this aperiodic behavior, extreme events are not predictable on long time scales, as found in Sec.~\ref{revlyapresults}.
The rarity of the events is related to the ``width'' of the opening of the channel and depends upon the parameters of the System~\cite{Karnatak2014}.

\begin{figure}
	\includegraphics{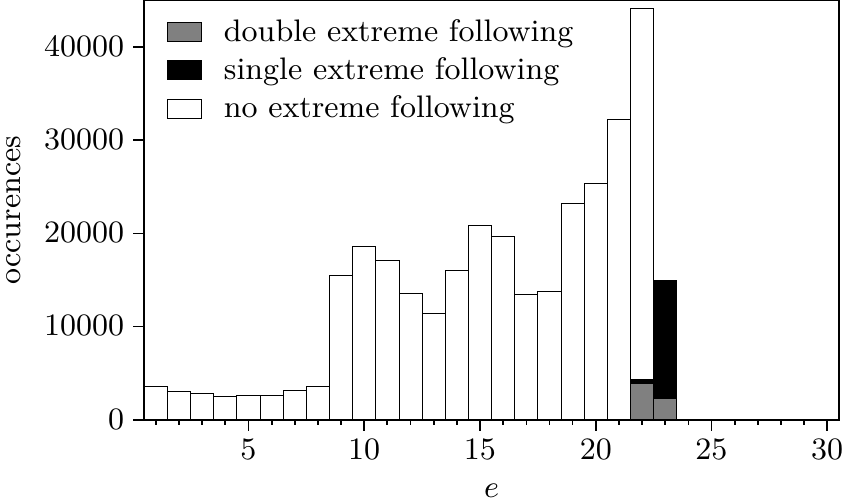}
	\caption{(Color online) Histogram (with non-overlapping bars) of $e$ for the proto-events from $2\cdot10^8$ time units of observation, separated into those that are followed by no extreme event or a single or double extreme event, respectively.
	There are no more than 200 proto-events for all $e$ between 23 and~60.
	Also, we observed no extreme event of either kind that was not preceded by a proto-event.
	}
	\label{fig:spike_statistics}
\end{figure}

\section{How an extreme event is generated in System~\ref{manyunits}}\label{genmany}

In this section we investigate how the emergence of extreme events in System~\ref{manyunits} depends on the proto-events, which we define for this purpose as local maxima of $e$ that are no extreme events, $e\kl{t}\defi\abs{\klg{i\middle|x_i\kl{t}>0.6}}$ being the number of excited units.
In Fig.~\ref{fig:spike_statistics} we show a histogram of the $e$ for these proto-events.
We observe that proto-events with $e<22$ are never followed by extreme events, 10\,\% of proto-events with $e=22$ are followed by extreme events and almost all proto-events with $e=23$ or higher are followed by extreme events.
Furthermore, we observe that there are almost no proto-events with $e>23$ and all extreme events are preceded by proto-events.

For a system of $n=1001$ completely coupled units with comparable parameters, we make similar observations:
Proto-events with $e<223$ are never followed by an extreme event, almost all proto-events with $e\geq224$ are followed by an extreme event, there are almost no proto-events with $e>226$ and all extreme events are preceded by proto-events.

We therefore conclude that extreme events emerge from a special case of the rather normal proto-events.
For this to happen, it is necessary that a certain ``critical mass'' of units becomes excited in the proto-event.
Moreover, if such a critical mass of units becomes excited, an extreme event is likely to happen.
Thus, proto-events with a critical mass of excited units can be considered as precursors to extreme events.
This is in accordance with our previous observations that indicate that a generating mechanism comes into action only about when the proto-event begins, i.e., about 50~time units before the extreme event (see Sec.~\ref{revlyapresults}).

\section{Conclusions}\label{conclusions}
We reported on three deterministic systems which are composed of diffusively coupled, inhomogeneous FitzHugh--Nagumo units and which are capable of generating extreme events.
Those systems are of increasing complexity, from more simple systems, on which we performed most of our analyses, to a complex network of 10000 units, which indicates a certain robustness of the phenomenon regarding the coupling topology.
It remains to be investigated whether comparable phenomena can be observed on other coupling topologies, e.g., with a hub structure.
The occurrence of the extreme events, though self-generated by a deterministic system without the influence of noise or any change of control parameters, does not exhibit signs of determinism:
the inter-event intervals are distributed nearly exponentially and we found no indicators for long-term predictability.

For the extreme events, we observed for all three models that first a portion of units becomes excited, which then recruits the rest of the units via the diffusive coupling, such that all or almost all units become excited, which constitutes the extreme event.
Despite these commonalities, we observed differences between the systems in the generation of extreme events:
While for a system of two units, whenever one unit becomes excited, it recruits the other and an extreme event is generated; in a system of 101~completely coupled units, it happens rather often that a portion of units becomes excited and only if their number exceeds a certain ``critical mass'', an extreme event is generated.
In the small-world system, the initial excitation spreads on the underlying lattice, which eventually leads to the extreme event.
Taking a different point of view, we found for a system of two units that the backbone of the extreme-event generation is a channel-like structure in state space that is entered by the system rarely and aperiodically because of its chaoticity and that when entered leads to a long excursion in state space, which constitutes the extreme event.
Whether a similar mechanism is at work for our more complex systems, cannot be determined, since computations of stable and unstable manifolds are not feasible in such high-dimensional state spaces.

Dynamics similar to the ones analyzed here have been found for other systems \cite{Marino2004, Zaks2005, Pisarchik2011}, however with different mechanisms leading to the extreme events.
In Ref.~\cite{Marino2004}, rare high-amplitude pulses were observed in the output of a semiconductor optical amplifier, whose phase-space structure is equivalent to that of a FitzHugh--Nagumo-type oscillator.
These pulses were interpreted as excitations being caused by the unavoidable experimental noise.
In Ref.~\cite{Zaks2005}, similar trajectories were observed in globally coupled networks of FitzHugh--Nagumo units that are subject to noise.
A noise-induced intermittent occurence of large events appearing in between long stretches of irregular small-scale oscillations will turn into regular occurences with shorter and shorter inter-event intervals as the noise strength is increased.
In Ref.~\cite{Pisarchik2011}, the phenomenon of the emergence of rare large pulses of light intensity in a pumped laser relies on the coexistence of different attractors in the system for a given set of parameters.
Again the noise applied to the pump current is responsible for the jumps between the coexisting attractors which is manifested as the high-amplitude pulse.
While in these studies, the emergence of peaks with very large amplitude is noise-induced, the formation of extreme events investigated here is entirely based on deterministic dynamics.

In Ref.~\cite{Reinoso2013} the dynamics of a semiconductor laser with optical feedback was investigated and the authors also reported on the formation of extreme pulses in a deterministic model.
The emergence of extreme pulses is closely related to the expansion of the attractor under a variation of the feedback strength.
In two coupled lasers in a master--slave configuration, large intensity pulses---called optical rogue waves in laser systems---have been observed experimentally~\cite{Bonatto2011}, and a theoretical investigation revealed that they occur in the vicinity of a crisis~\cite{ZamoraMunt2013}.
The corresponding distribution of the inter-event intervals is also exponential, which hints to commonalities with our generating mechanism that need to be further explored.

There are other model systems, whose dynamics exhibits extreme pulses localized in space and time, such as the complex Ginzburg--Landau equation \cite{Kim2003, Nagy2007} and the nonlinear Schr\"odinger equation~\cite{Chabchoub2011}.
The mechanism of their appearance is very distinct from the one reported here since it includes only next-neighbor spatial interactions on a lattice.
Pulse-coupled oscillators with a complex coupling topology were shown to exhibit extreme events of synchrony, which emerge spontaneusly from an asynchronous chaotic behavior~\cite{Rothkegel2011}, similar as observed here.

From these considerations we can conclude that particularly optical rogue waves and the extreme events found in systems of coupled oscillators seem to have more in common than it seems at first glance.
Hence, future studies could address the question to what extent there are commonalities and differences in the extreme-event-generating mechanisms in these systems.

\section*{Acknowledgments}
We authors would like to thank Stephan Bialonski, Jan Freund, Holger Kantz, J\"urgen Kurths, Cristina Masoller, Jos\'e Rios Leite, Alexander Rothkegel, and Jorge Tredicce for interesting discussions.
We are grateful to Stephan Bialonski and Alexander Rothkegel for critical comments on earlier versions of the manuscript.
This work was supported by the Volkswagen Foundation (Grant~Nos. 85388 and 85392).
U.F.~would like to thank R.~Roy and his group for their hospitality and the Burgers Program for Fluid Dynamics of the University of Maryland for financial support.

\appendix
\section{Manifold approximation} \label{sec:app1}
The eigenvalues of the Jacobian at the origin are $\lambda_{1,2}=0.00041\pm 0.09683 \, i$,
$\lambda_3=-0.016225$, and $\lambda_4=-0.082989$.
The associated saddle quantity is $\sigma = 2\, \text{Re}\kl{\lambda_{1,2}}+ \lambda_3+ \lambda_4 \approx -0.098394$.
These eigenvalues suggest that the stable and the unstable manifolds are both two-dimensional for this case.
The stable manifold can be approximated~\cite{Yanhui2006} using the expressions
	\begin{align}\nonumber
		x_1\kl{t}&=\sum_{l}\sum_{m=0}^{l}\rho_{m,l-m}\exp\kl{m\lambda_1 t+(l-m)\lambda_2 t},\\\nonumber
		y_1\kl{t}&=\sum_{l}\sum_{m=0}^{l}\varsigma_{m,l-m}\exp\kl{m\lambda_1 t+(l-m)\lambda_2 t},\\\nonumber
		x_2\kl{t}&=\sum_{l}\sum_{m=0}^{l}\tau_{m,l-m}\exp\kl{m\lambda_1 t+(l-m)\lambda_2 t},\\
		y_2\kl{t}&=\sum_{l}\sum_{m=0}^{l}\chi_{m,l-m}\exp\kl{m\lambda_1 t+(l-m)\lambda_2 t},
		\label{eq:stableman}
	\end{align}
where $\rho_{m,l-m}$, $\varsigma_{m,l-m}$, $\tau_{m,l-m}$, $\chi_{m,l-m}$ are undetermined coefficients.
$l\in[1,2,3, \ldots]$ can be considered as the order of the expansion.
Similarly, the unstable manifold can be approximated from
	\begin{align}\nonumber
		x_1\kl{t}&=\sum_{l}\sum_{m=0}^{l}\rho_{m,l-m}'\exp\kl{m\lambda_3 t+(l-m)\lambda_4 t},\\\nonumber
		y_1\kl{t}&=\sum_{l}\sum_{m=0}^{l}\varsigma_{m,l-m}'\exp\kl{m\lambda_3 t+(l-m)\lambda_4 t},\\\nonumber
		x_2\kl{t}&=\sum_{l}\sum_{m=0}^{l}\tau_{m,l-m}'\exp\kl{m\lambda_3 t+(l-m)\lambda_4 t},\\
		y_2\kl{t}&=\sum_{l}\sum_{m=0}^{l}\chi_{m,l-m}'\exp\kl{m\lambda_3 t+(l-m)\lambda_4 t},
		\label{eq:unstableman}
	\end{align}
where $\rho_{m,l-m}'$, $\varsigma_{m,l-m}'$, $\tau_{m,l-m}'$, and $\chi_{m,l-m}'$ are once again the undetermined coefficients with $l$ being the order of the expansion.
Substituting the expansions from Eqs.~\ref{eq:stableman} and~\ref{eq:unstableman} into the system Eq.~\ref{eq:systems} and matching the terms of same order of the exponential on both sides of the resultant equation, these coefficients can be determined.
The zeroth-order coefficients vanish; $\kl{\rho_{0,0}, \varsigma_{0,0}, \tau_{0,0}, \chi_{0,0}} = \kl{\rho_{0,0}', \varsigma_{0,0}', \tau_{0,0}', \chi_{0,0}'} = \kl{0,0,0,0}$ as the manifolds pass through the origin.
The first-order coefficients $(l=1)$ can be approximated by using the eigenvectors of the eigenvalues of the Jacobian at the origin:
	  \begin{align}\nonumber
	    \rho_{1,0}&=\mu_1v_{1,1}; \varsigma_{1,0}=\mu_1v_{1,2}; \tau_{1,0}=\mu_1v_{1,3}; \chi_{1,0}=\mu_1v_{1,4}\\\nonumber
        \rho_{0,1}&=\mu_2v_{2,1}; \varsigma_{0,1}=\mu_2v_{2,2}; \tau_{0,1}=\mu_2v_{2,3}; \chi_{0,1}=\mu_2v_{2,4}\\\nonumber
        \rho_{1,0}'&=\mu_3v_{3,1}; \varsigma_{1,0}'=\mu_3v_{3,2}; \tau_{1,0}'=\mu_3v_{3,3}; \chi_{1,0}'=\mu_3v_{3,4}\\\nonumber
        \rho_{0,1}'&=\mu_4v_{4,1}; \varsigma_{0,1}'=\mu_4v_{4,2}; \tau_{0,1}'=\mu_4v_{4,3}; \chi_{0,1}'=\mu_4v_{4,4},\\
	    \label{eq:coeffvals}
	  \end{align}
where $v_{i,j}$ is the $j$-th component of the eigenvector corresponding to the $i$-th eigenvalue.
Constants $\mu_i$ are picked to keep the series expansions in Eqs.~\ref{eq:stableman} and~\ref{eq:unstableman} convergent.
The undetermined higher-order coefficients can be calculated simultaneously and recursively for $l=1,2,\ldots$.
For our calculations, we consider $\mu_i=\mu=0.01$, and use the expansions in Eqs.~\ref{eq:stableman} and~\ref{eq:unstableman} up to $l=5$.
The estimation of the coefficients was performed using Maple~\cite{Maple10}.

\end{document}